\documentclass{iau} 
\usepackage{graphicx}

\title[JD 11.~~Galactoseismology] 
{Galactoseismology in the GAIA Era}

\author[Sukanya Chakrabarti]   
{Sukanya Chakrabarti$^1$}

\affiliation{$^1$ 84 Lomb Memorial Drive, \\ School of Physics and Astronomy,
RIT, Rochester NY 14623 \\ email: {chakrabarti@astro.rit.edu} \\}

\pubyear{2016}
\volume{321}  
\jname{Formation and Evolution of Galaxy Outskirts}
\editors{Gil de Paz, A., Knapen, J. H. and Lee, J. C., eds.}
\begin{document}

\maketitle

\begin{abstract}

The GAIA satellite will provide unprecedented phase-space information for our Galaxy and enable a new era of Galactic dynamics.   We may soon see successful realizations of Galactoseismology, i.e., inferring the characteristics of the Galactic potential and sub-structure from a dynamical analysis of observed perturbations in the gas or stellar disk of the Milky Way.  Here, we argue that to maximally take advantage of the GAIA data and other complementary surveys, it is necessary to build comprehensive models for both the stars and the gas.  We outline several key morphological puzzles of the Galactic disk and proposed solutions that may soon be tested.

\end{abstract}

\firstsection 
\section{Introduction}

Connecting the puzzling disturbances in both the gas and stellar disk of the Milky Way (MW) with the dark matter distribution of our Galaxy and its dwarf companions may become possible in the GAIA era (Perryman et al. 2001). GAIA will provide parallaxes and proper motions for a billion stars down to $V \sim 20$ (de Bruijne et al. 2014) and radial velocities for stars with $V <15$.  By now, a plethora of stellar tidal streams have been discovered, including the Sagittarius (Sgr) tidal stream (Ibata et al. 1997), the Monoceros stream (Newberg et al. 2002), and many others (Belokurov et al. 2006).  A number of authors have attempted to infer the Galactic potential by modeling stellar tidal streams (e.g. Johnston et al. 1999), but the limitations of determining accurate phase space information for the stream and simplistic modeling (for example static halos) have led to large uncertainties in the reconstruction of the Galactic potential.  More recently, observations of an asymmetry in the number density and bulk velocity of solar neighborhood stars have been interpreted as arising from a dark sub-halo or dwarf galaxy passing through the Galactic disk, exciting vertical waves (Widrow et al. 2012; Carlin et al. 2013; Xu et al. 2015). This corroborates a similar previous suggestion that the disturbances in the outer HI disk of our Galaxy may be due to a massive, perturbing satellite (Chakrabarti \& Blitz 2009; henceforth CB09).  There is some evidence now for this predicted satellite, which may mark the first success of Galactoseismology (Chakrabarti et al. 2016).

Galaxy outskirts hold particularly important clues to the past galactic accretion history and dynamical impacts.  Extended HI disks reach to several times the optical radius (Walter et al. 2008), presenting the largest possible cross-section for interaction with sub-halos at large distances (where theoretical models \emph{expect} them to be, e.g. Springel et al. 2008).  The gas disk of our Galaxy manifests large planar disturbances and is warped (Levine, Blitz \& Heiles 2006).   Chakrabarti \& Blitz (2009; 2011) found that these puzzling planar disturbances in the gas disk of our Galaxy could be reproduced by an interaction with a sub-halo with a mass one-hundredth that of the Milky Way, with a pericenter distance of $\sim$ 7 kpc, which is currently at $\sim$ 90 kpc.  This interaction also produces structures in the stellar disk that are similar to the Monoceros stream at present day.  Chakrabarti et al. (2015) found an excess of faint variables at $l \sim 333 ^\circ$, and Chakrabarti et al. (2016) obtained spectroscopic observations of three Cepheid candidates that are part of this excess.  The average radial velocities of these stars is $\sim$ 163 km/s, which is large and distinct from the stellar disk of the Galaxy (which in the fourth quadrant is negative).  Using the period-luminosity relations for Type I Cepheids, we obtained an average distance of 73 kpc for these stars (Chakrabarti et al. 2016). 

Tidal interactions remain manifest in the stellar disk for many crossing times, but the gas is collisional and disturbances in the gas disk dissipate on the order of a dynamical time.  Therefore, an analysis of disturbances in the gas disk can provide a constraint on the time of encounter (Chakrabarti et al. 2011).  Ultimately, a joint analysis of the gas (a cold, responsive, dissipative component that is extended such as the HI disk) \emph{and} the stars (that retain memory of the encounter for many crossing times) holds the most promise for unearthing clues about recent \emph{and} past encounters.  

\begin{figure}[h]
\begin{center}
 \includegraphics[width=3.in]{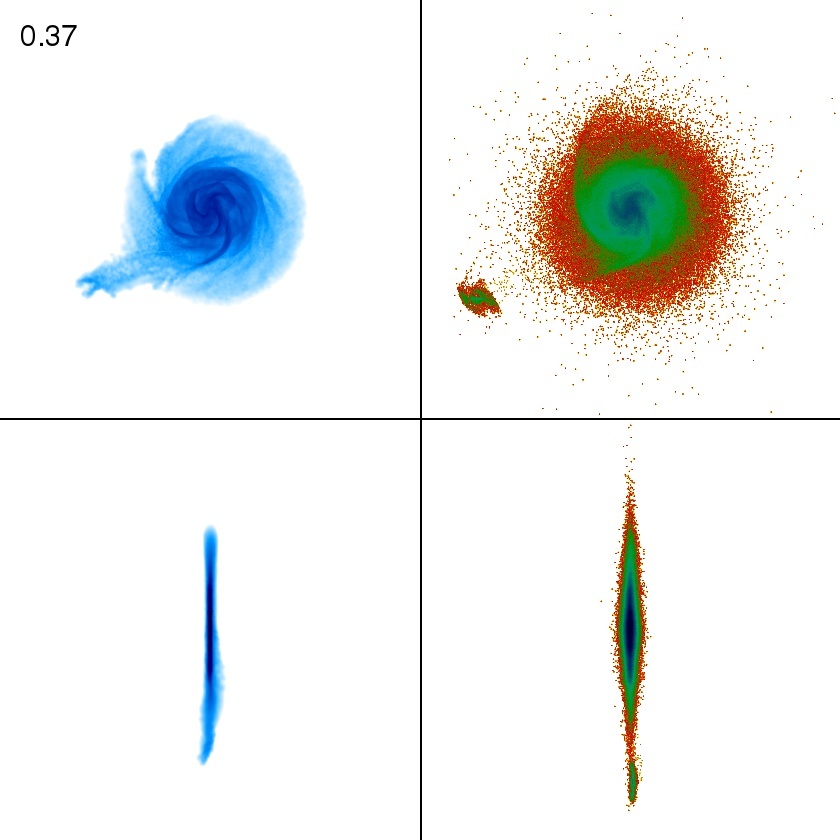} 
 \caption{(Left) Gas density image of the Milky Way after undergoing an encounter with a $\sim$ 1:100 mass ratio perturber, (Right) an image of the stellar density distribution.  From Chakrabarti \& Blitz (2009). }
   \label{fig1}
\end{center}
\end{figure}

\section{Overview}

Extended HI disks of local spirals have low sound speeds compared to their rotation velocity, and so are extremely sensitive to gravitational disturbances.   Furthermore, in the outskirts, atomic hydrogen traces the bulk of the ISM (Bigiel et al. 2010). Therefore, the outskirts of galaxies are less subject to the effects of feedback from supernovae and star formation that complicate the ISM structure (and the modeling thereof) in the inner regions of galaxies (Christensen et al. 2013).  Using the sensitivity of gaseous disks to disturbances, we constrained the mass and current
radial distance of galactic satellites (Chakrabarti et al. 2011; CB11; CB09) and its azimuth to zeroth order by finding
the best-fit to the low-order Fourier modes (i.e., low m modes that trace large-scale structures, $>$ kpc-
scale, in the disk) of the projected gas surface density of an observed galaxy.  We tested our ability to characterize the galactic satellites of spirals with optically visible companions, namely, M51 and NGC 1512, which span the range from having a very low mass companion ($\sim$ 1:100 mass ratio) to a fairly massive companion ($\sim$ 1:3 mass ratio). We accurately recover the masses and relative positions of the satellites in both these systems (Chakrabarti et al. 2011).  To facilitate a statistical study, we developed a simplified numerical approach along with a semi-analytic method to study the excitation of disturbances in galactic disks by passing satellites, and derived a simple scaling relation between the mass of the satellite and the sum of the Fourier modes (Chang \& Chakrabarti 2011).  We later extended this method to also constrain the dark matter density profile of spiral galaxies (Chakrabarti 2013).

\section{Future}

Of particular interest now with the advent of GAIA, is if we can detect the kinematical signature of this interaction in the stars that it perturbed at pericenter.  If the stars for which radial velocities were obtained by Chakrabarti et al. (2016) are indeed part of the dwarf galaxy predicted by CB09, then such a detection would enable a constraint on the orbit and angular momentum of this dwarf galaxy.  Price-Whelan et al. (2013) have noted that the GAIA data can be complemented by measuring RR Lyrae stars in the mid-infrared, which would allow for distances accurate to 2 \% out to $\sim$ 30 kpc, i.e., this would give accurate distances for the outer HI disk.  The puzzles of the Milky Way disk -- the large ripples in the gas disk, the many stellar streams, and the vertical waves in the Galactic disk, need to be studied comprehensively.   Only for our Galaxy, can we connect the dots between the orbits, the dynamical evolution of the satellites, the disk structure, and its place in the broader context of galaxy formation.  A joint analysis of the data from GAIA and HI surveys of the Milky Way should enable this effort.

\acknowledgments

SC gratefully acknowledges support from NSF grant 1517488.

\end{document}